\documentclass[journal]{IEEEtran}

\usepackage{ragged2e}
\usepackage{floatrow}
\usepackage{amsthm}
\usepackage[utf8]{inputenc}
\usepackage[english]{babel}
\usepackage{multicol}
\usepackage{epsfig}
\usepackage{epstopdf}
\usepackage{float}
\usepackage{perpage} 
\usepackage{array}
\usepackage{graphicx}
\usepackage{amsfonts}
\usepackage{comment}
\usepackage{subcaption} 
\usepackage{algorithm}
\usepackage{algpseudocode} 
\MakeSorted{figure}
\MakeSorted{table}
\usepackage{bbm}
\usepackage{bm,xstring}
\usepackage{xcolor}
\usepackage{multirow}
\usepackage{soul}
\usepackage{cite}
\usepackage{array} 
\usepackage[utf8]{inputenc}
\usepackage{amsmath}
\usepackage{amssymb}
\theoremstyle{remark}

\newtheorem{lemma}{Lemma}

\usepackage{tabularx}
\usepackage{pifont}        

\usepackage{enumitem}
\usepackage{algpseudocode}  
\usepackage{hyperref}

\usepackage{cite}

\usepackage[font=footnotesize,labelfont=bf, figurename=Fig.]{caption} 
\usepackage{subcaption}
\usepackage{fix-cm}
\usepackage{array, makecell} 

\begin{document}

\title{Generative AI-Driven Phase Control for RIS-Aided Cell-Free Massive MIMO Systems}
\author{
Kalpesh K. Patel,  Malay Chakraborty, Ekant Sharma and Sandeep Kumar Singh  
\thanks{The authors are with the Department of Electronics and Communication Engineering, Indian Institute of Technology, Roorkee 247667, India (e-mail: \{kalpeshkumar\_p, malay\_c, ekant, sandeep.singh\}@ece.iitr.ac.in).}
\vspace{-15pt} }
\maketitle
\begin{abstract}
This work investigates a generative artificial intelligence (GenAI) model to optimize the reconfigurable intelligent surface (RIS) phase shifts in RIS-aided cell-free massive multiple-input multiple-output (mMIMO) systems under practical constraints, including imperfect channel state information (CSI) and spatial correlation. We propose two GenAI based approaches, generative conditional diffusion model (GCDM) and generative conditional diffusion implicit model (GCDIM), leveraging the diffusion model conditioned on dynamic CSI to maximize the sum spectral efficiency (SE) of the system. To benchmark performance, we compare the proposed GenAI based approaches against an expert algorithm, traditionally known for achieving near-optimal solutions at the cost of computational efficiency. The simulation results demonstrate that GCDM matches the sum SE achieved by the expert algorithm while significantly reducing the computational overhead. Furthermore, GCDIM achieves a comparable sum SE with an additional $98\%$ reduction in computation time, underscoring its potential for efficient phase optimization in RIS-aided cell-free mMIMO systems. 

\end{abstract}

\vspace{-5pt}

\IEEEpeerreviewmaketitle
\vspace{-10pt}
\section{Introduction}

The rapid growth of future 6G networks demands high capacity, low latency, and widespread coverage. Cell-free massive multiple-input multiple-output (mMIMO) meets these requirements by using distributed access points (APs) that collaboratively serve the users~\cite{Lit_cellfree_2017}. However, the need for a large number of active APs significantly increases both deployment costs and overall power consumption. To overcome this, researchers adopt reconfigurable intelligent surface (RIS) as an energy-efficient solution, where optimized passive elements improve system performance at low cost and reduce power usage~\cite{Chien2022RISCellFree}.  
The deployment of RIS in wireless systems, although promising, leads to intricate optimization problems that must be addressed. The authors in~\cite{Lit_RIS_2024_conventional3} presented a comprehensive survey, reviewing various optimization techniques for RIS-aided wireless networks, encompassing model-based methods, heuristic algorithms, and machine learning approaches. In ~\cite{[CR1]}, the authors formulated an energy efficiency maximization problem based on instantaneous SINR and proposed a block coordinate descent based algorithm to optimize the RIS reflection coefficients in a RIS-aided cell-free system. The authors in~\cite{ourpaper} introduced a meta-heuristic based particle swarm optimization technique to optimize the phase shifts in a RIS-assisted cell-free system. To this end, studies in \cite{Lit_RIS_2024_conventional3,[CR1],ourpaper} have adopted model-based or heuristic approaches, often proving inefficient in terms of computational efficiency.

Recently, generative artificial intelligence (GenAI) has accelerated AI advancements by introducing powerful generative models such as variational autoencoders (VAEs), generative adversarial networks (GANs), and diffusion models. Diffusion models outperform GANs by avoiding mode collapse and training instability, and also surpass VAEs by producing high-fidelity outputs~\cite{Lit_Diffusion_2024_CDDM}. 
A few studies have recently explored the application of diffusion models to various problems in cell-free mMIMO systems. For instance, the authors in \cite{Lit_Diffusion_2024_rate} optimize power-splitting factors and power-control coefficients for rate-splitting in cell-free mMIMO systems, while in \cite{Lit_Diffusion_2024_hardware}, enhance downlink signal reconstruction by mitigating hardware impairments and interference. Reference~\cite{lit_genflownet_2024} employed a generative flow network paradigm for RIS phase optimization in a single-input single-output system. The authors in \cite{Lit_RIS_DRL_2024} proposed a multi-agent reinforcement learning approach for RIS phase shift optimization in cell-free mMIMO systems. However, such methods often face scalability challenges due to the complex interactions among multiple agents. Due to the stable training, strong mode coverage, and high robustness, we choose diffusion models as the preferred approach for phase optimization in RIS-aided cell-free systems. \textit{To the best of our knowledge, no existing work has employed diffusion based GenAI approaches to optimize the RIS phase coefficients in RIS-aided cell-free systems.} The \textbf{key contributions} of this paper are as follows:

\begin{itemize}\addtolength{\leftskip}{-10pt}
    \item We propose a GenAI based framework for optimizing the RIS phase shifts to maximize the sum spectral efficiency (SE) in RIS-aided cell-free mMIMO systems while accounting for practical constraints such as imperfect channel state information (CSI) and spatial correlation. We develop two GenAI based approaches, namely, generative conditional diffusion model (GCDM) and generative
conditional diffusion implicit model (GCDIM), to achieve optimal performance while significantly reducing the time complexity. We condition the diffusion process with dynamic CSI, enabling the model to generalize across diverse channel states to capture the time-varying nature of the wireless environment. 

    \item Numerical results validate the effectiveness of the proposed GCDM and GCDIM approaches through extensive simulations, demonstrating that GCDM matches the sum SE achieved by the expert algorithm while significantly reducing the computational overhead. Furthermore, the GCDIM offers even greater computational efficiency with minimal performance loss. 
    
\end{itemize}

\vspace{-10pt}
\section{{System Model and Problem Formulation}}
We consider an RIS-assisted cell-free mMIMO system, as depicted in Fig.~\ref{fig_Gen_ai}, where $M$ APs serve $K$ single-antenna users with the aid of an RIS comprising $N$ passive reflecting elements. The network supports both direct and reflected communication links, thereby enhancing signal diversity. All nodes operate under time-division duplex mode. The coherence interval is denoted as $\tau_c = \tau_p + \tau_d$, where $\tau_p$ and $\tau_d$ represent the durations for uplink training and downlink data transmission, respectively.

The channels from the $m$th AP to the $r$th RIS and the $r$th RIS to the $k$th user are represented by $\mathbf{g}_{mr} \in \mathbb{C}^{N \times 1}$ and $\mathbf{h}_{rk}^{H} \in \mathbb{C}^{1 \times N}$, respectively. These channels are modeled as
\begin{align}
    \mathbf{g}_{mr} &= \mathbf{R}_{mr}^{1/2} \bar{\mathbf{g}}_{mr}, \quad
    \mathbf{h}_{rk} = \mathbf{R}_{rk}^{1/2} \bar{\mathbf{h}}_{rk},
\end{align}
where $\bar{\mathbf{g}}_{mr}, \bar{\mathbf{h}}_{rk} \sim \mathcal{CN}(\mathbf{0}, \mathbf{I}_L)$ denote small-scale fading components, and $\mathbf{R}_{mr}, \mathbf{R}_{rk} \in \mathbb{C}^{N \times N}$ are the covariance matrices. We model these matrices as $\mathbf{R}_{mr} = \beta_{mr} A \mathbf{R}$ and $\mathbf{R}_{rk} = \beta_{rk} A \mathbf{R}$, where $\beta_{mr}$ and $\beta_{rk}$ are large-scale fading coefficients, $A$ is the area of each RIS element, and $\mathbf{R}$ is the spatial correlation matrix \cite{Chien2022RISCellFree}. The direct channel between AP $m$ and user $k$ is given as $l_{mk} = \sqrt{\beta_{mk}} \bar{l}_{mk}$, where $\bar{l}_{mk} \sim \mathcal{CN}(0,1)$ denotes the small-scale fading coefficients and $\beta_{mk}$ represents the large-scale fading coefficients. The aggregated downlink channel from AP $m$ to user $k$ is expressed~as
\begin{align}\label{virtual_channel}
    u_{mk} = l_{mk} + \mathbf{h}_{rk}^{H} \boldsymbol{\Theta}_r \mathbf{g}_{mr},
\end{align}
where $\boldsymbol{\Theta}_r = \text{diag}(e^{j\theta_1},\ldots,e^{j\theta_N}) \in \mathbb{C}^{N \times N}$ is the RIS phase shift matrix with $\theta_i \sim \mathcal{U}[-\pi, \pi)$.

During uplink training, each of the $K$ users transmits an orthogonal pilot sequence $\sqrt{\tau_p} \boldsymbol{\varphi}_k \in \mathbb{C}^{\tau_p \times 1}$ such that $\|\boldsymbol{\varphi}_k\|^2 = 1$ and $\boldsymbol{\varphi}_k^H \boldsymbol{\varphi}_{k'} = 0$ for $k \neq k'$. Assuming $\tau_p \geq K$, the linear minimum mean square error (LMMSE) estimate of the channel $u_{mk}$ is given by \cite{Chien2022RISCellFree}
\begin{align}
    \hat{u}_{mk} = c_{mk} \left(\sqrt{\tau_p p_p} u_{mk} + \boldsymbol{\varphi}_k^H \mathbf{w}_{pm} \right),
\end{align}
where $p_p$ is the normalized pilot power and $\mathbf{w}_{pm} \sim \mathcal{CN}(\mathbf{0}, \sigma_n^2 \mathbf{I}_{\tau_p})$ denotes the additive white Gaussian noise (AWGN) at the $m$th AP. The scalar $c_{mk}$ is defined as
\begin{align}
    c_{mk} = \frac{\sqrt{\tau_p p_p} \delta_{mk}}{\tau_p p_p \delta_{mk} + 1},
\end{align}
where $\delta_{mk} = \beta_{mk} + \text{Tr}(\boldsymbol{\Theta}_r \mathbf{R}_{mr} \boldsymbol{\Theta}_r^H \mathbf{R}_{rk})$ is the variance of the aggregated channel. The estimation error $\varepsilon_{mk} \triangleq u_{mk} - \hat{u}_{mk}$ is uncorrelated with $\hat{u}_{mk}$ and has zero mean and variance $\mathbb{E}[|\varepsilon_{mk}|^2] =\delta_{mk} - \gamma_{mk}$, with $\gamma_{mk} = \mathbb{E}[|\hat{u}_{mk}|^2] = \sqrt{\tau_p p_p} \delta_{mk}c_{mk}$ is the estimated channel variance.

In the downlink data transmission phase, each AP utilizes conjugate beamforming based on statistical CSI to transmit the information signal~\cite{Lit_cellfree_2017}. The signal transmitted by AP $m$ is expressed as \vspace{-5pt}
\begin{align}
    x_m = \sum_{k=1}^{K} \sqrt{\eta_{mk} \rho_d} \hat{u}_{mk}^* s_k,
\end{align} 
where $s_k \sim \mathcal{CN}(0,1)$ is the information symbol for user $k$, $\rho_d$ is the maximum transmit power, and $\eta_{mk}$ is the power control coefficient satisfying $\sum_{k=1}^K \eta_{mk} \gamma_{mk} \leq 1$. The received signal at user $k$ can be expressed as 
\begin{align}\label{rx_components}
{r}_{k} &=\sum\limits_{m = 1}^M u_{mk}x_{m} + \upsilon_{k}=\underbrace{\sum\limits_{m = 1}^M \!\!\sqrt{\eta_{mk}\rho_{d}}u_{mk} {\hat{u}}_{mk}^{*}}_{\text{Desired Signal}} s_{k} \nonumber\\[-10pt]&+\underbrace{\sum\limits_{\substack{k' \neq k}}^K\sum\limits_{m = 1}^M \!\!\sqrt{\eta_{mk'}\rho_{d}} u_{mk}{\hat{u}}_{mk'}^{*}}_{\text{User Interference}} s_{k'}+\underbrace{\upsilon_{k}}_{\text{AWGN}},
\end{align}
where $\upsilon_k \sim \mathcal{CN}(0,\sigma_n^2)$ is the AWGN at the $k$th user.
\vspace{-10pt}
\subsection{Achievable Spectral Efficiency}
In this section, we derive a closed-form lower bound for the achievable sum-SE. Utilizing the use-and-then-forget bounding technique~\cite{Lit_cellfree_2017}, the received signal at the $k$th user in \eqref{rx_components} can be reformulated as \eqref{rk_uatf} (shown at the top of the next page) to express i) the desired signal ($\text{D}_k$), ii) beamforming uncertainty ($\text{BU}_k$), iii) multi-user interference ($\text{MUI}_k$), and iv) AWGN.

\begin{figure*}
    \begin{align}\label{rk_uatf}
    {r}_{k}\!
    =\!\underbrace{\mathbb{E}\!\left\{\!\sum\limits_{m = 1}^M \!\!\sqrt{\eta_{mk}\rho_{d}}u_{mk}{\hat{u}}_{mk}^{*}\right\}}_{\text{D}_{k}} \!\!s_{k}\!+\!\underbrace{\left(\sum\limits_{m = 1}^M \!\!\sqrt{\eta_{mk}\rho_{d}}(u_{mk}{\hat{u}}_{mk}^{*}\!\!-\!\mathbb{E}\left\{u_{mk}{\hat{u}}_{mk}^{*}\right\})\right)}_{\text{BU}_{k}}\!\!s_{k} \!+\!\sum\limits_{\substack{k' \neq k}}^K\underbrace{\!\sum\limits_{m = 1}^M\!\!\! \sqrt{\eta_{mk'}\rho_{d}} u_{mk}{\hat{u}}_{mk'}^{*}}_{\text{MUI}_{k}} s_{k'}\!+\!\!\underbrace{\upsilon_{k}}_{\text{AWGN}}\!\!.
\end{align} \hrule \vspace{-10pt}
 \end{figure*}

\begin{lemma}
For the RIS-assisted cell-free mMIMO system employing LMMSE-based conjugate beamforming, the closed-form expression for the achievable SE is given by
\begin{align}
    \mathcal{R}_\text{sum} = \sum_{k=1}^{K} \mathcal{R}_{k} = \left(\frac{\tau_d}{\tau_c}\right) \sum_{k=1}^{K} \log_2\left(1 + \Delta_k\right),
\end{align}
where the effective signal-to-interference-plus-noise ratio (SINR) for user \( k \), denoted \( \Delta_k \), is defined as
\begin{align} \label{Gamma_k}
 \Delta_k = \frac{|\text{D}_{k}|^2}{\mathbb{E}\{|\text{BU}_{k}|^2\} + \mathbb{E}\{|\text{MUI}_{k}|^2\} + \sigma_n^2}.
\end{align}
Solving each term in \eqref{Gamma_k}, the closed-form SINR expression at the $k$th user can be obtained as~\cite{Chien2022RISCellFree,ourpaper}
\begin{align}\label{Gamma_cf}
    \Delta_k = \frac{\left| \sum\limits_{m = 1}^M \sqrt{\eta_{mk} \rho_{d}}\,\gamma_{mk} \right|^2}
    {\rho_d \sum\limits_{m=1}^{M} \eta_{mk}\delta_{mk}\gamma_{mk} +
    \rho_d \sum\limits_{k' \ne k}^K \sum\limits_{m=1}^M \eta_{mk'}\delta_{mk}\gamma_{mk'} + \sigma_n^2}.\vspace{-2pt}
\end{align}
\end{lemma}

\vspace{-15pt}
\subsection{{Problem Formulation}} 
We now define the problem formulation for optimizing the downlink achievable SE for the RIS-aided cell-free mMIMO system. Our goal is to enhance the achievable SE by strategically designing the RIS phase shift vector  $\boldsymbol{\theta} = (\theta_1, \cdots, \theta_N)$. We formulate an optimization Problem \textbf{P1} aimed to maximize the achievable SE by optimizing $\boldsymbol{\theta}$ as follows 
\begin{subequations}\label{P1}
 \begin{align} \textbf{P1:}\quad
     \underset{\boldsymbol{\theta}}{\text{maximize}} \;\; & \left(\frac{\tau_d}{\tau_c}\right) \sum\limits_{k=1}^{K}\log_2\left(1+\Delta_{k}(\boldsymbol{\theta})\right)\!,\label{P1_a} \\
     \text{subject to}\;\;  
     &\theta_{n} \in [0,2\pi), \;\; n\in \{1,2,\dots,N\},\label{P1_b}   
 \end{align} 
 \end{subequations}
The constraint in \eqref{P1_b} defines the allowable range for the phase shifts applied by each RIS element. Problem \textbf{P1} is highly non-convex owing to the tight coupling of phase shifts across terms in the closed-form SINR expression in~\eqref{Gamma_cf}, making traditional convex optimization methods ineffective. Although meta heuristic approach like genetic algorithm~\cite{GA_1994} succeed in locating the optimal solution by preserving the best individuals throughout the evolutionary process, their high computational complexity limits practical deployment. Hence, we propose GenAI based approaches to efficiently optimize the RIS phase shifts for the above non-convex problem. 
\begin{figure}   
    \includegraphics[scale=0.42]{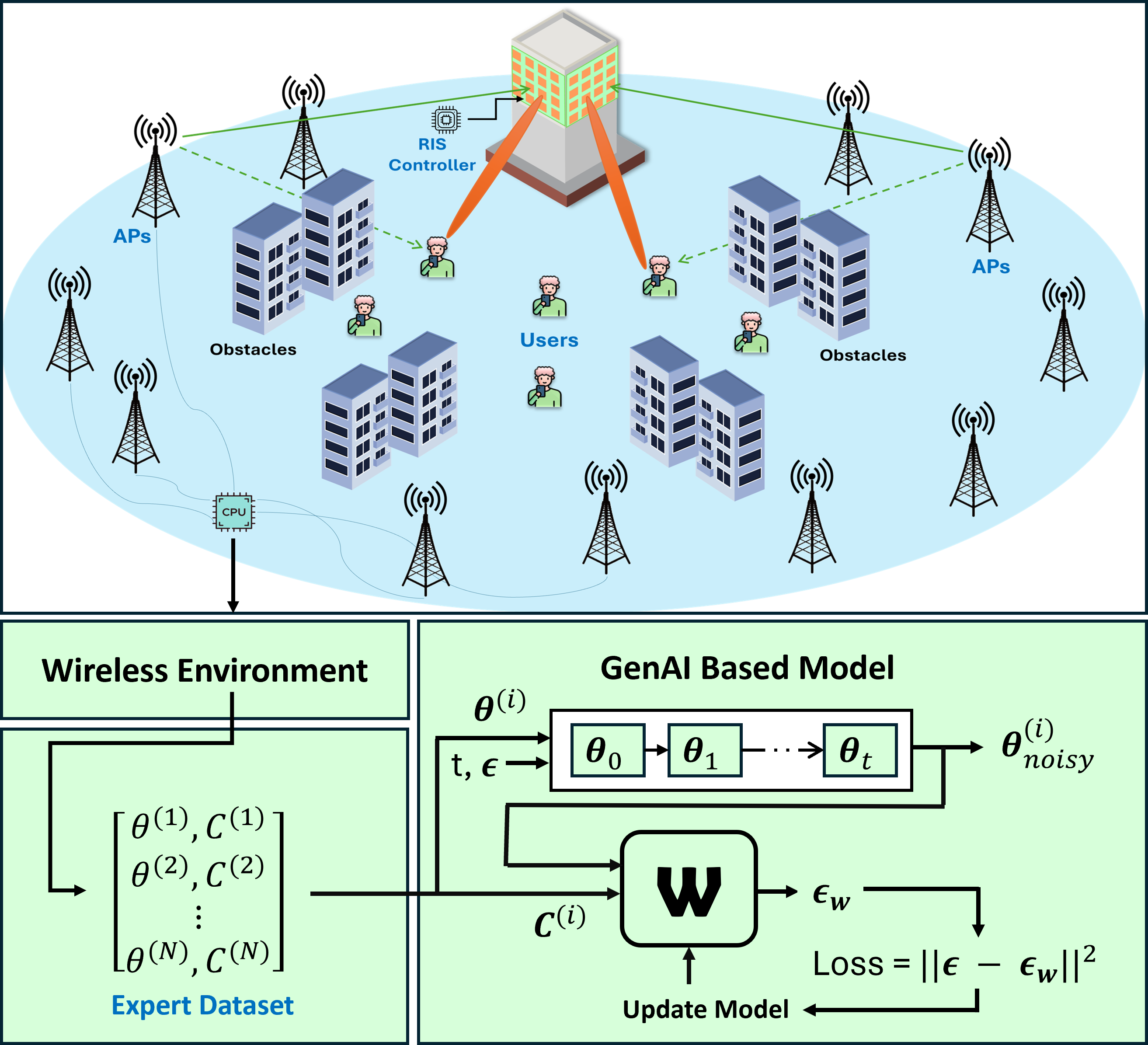}
    \caption{\small Proposed GenAI framework for RIS phase optimization. \label{fig_Gen_ai} } 
\end{figure} 

\vspace{-10pt}
\section{RIS Phase Optimization With GenAI}
We now propose two GenAI based approaches, namely GCDM and GCDIM, designed to optimize the RIS phase shifts in the considered RIS-aided cell-free mMIMO system. 

\vspace{-10pt}
\subsection{{Motivation of using the GenAI based approaches}}
The motivation for adopting GenAI based approaches arises from their strong capability to tackle complex, high-dimensional optimization problems~\cite{motivation_2024}. Unlike traditional discriminative models that rely on fixed input-output mappings and often perform poorly in dynamic, non-convex scenarios, GenAI models learn the joint distribution of inputs and outputs. This capability allows GenAI models to explore a broader solution space and generate diverse, high-quality results. By sampling from these distributions, they exhibit greater adaptability and robustness to unseen inputs, making them particularly effective for optimization tasks in wireless communication ~\cite{motivation_2024}.

Moreover, traditional optimization methods like exhaustive search can find optimal solutions but are computationally infeasible for problems with many optimization variables, i.e., number of RIS elements. Alternative approaches, like block coordinate descent (BCD) and successive convex approximation (SCA) provide suboptimal solutions involving intricate mathematical derivations~\cite{[CR1]}. However, GenAI based approaches not only outperform traditional methods (such as exhaustive search, BCD, SCA) but also offer superior scalability and real-time adaptability~\cite{Lit_Diffusion_2024_rate}. Therefore, we propose GCDM and GCDIM, two GenAI based approaches, to tackle the phase shift optimization problem in RIS-aided cell-free mMIMO systems.

\subsection{{GCDM Based RIS Phase Optimization}}
We introduce GCDM, a GenAI based approach that utilizes a diffusion model~\cite{Lit_GenAI_2020_DDPM} to learn the conditional distribution $p(\boldsymbol{\theta}_0|\boldsymbol{c})$ of optimal phase shift in RIS-aided cell-free mMIMO systems. Here,~$\boldsymbol{\theta}_{0} = ({\theta_1},\cdots\!,{\theta_{N}})$
denotes the RIS phase shift corresponding to channel state 
$\boldsymbol{c}$, which is generated by expert algorithm. Here, we take channel state $\boldsymbol{c}$ as set of the large scale fading coefficients, i.e., $\boldsymbol{c}=\{\beta_{mr}, \beta_{rk}, \beta_{mk}\},\forall m, k$. In our expert dataset  $\boldsymbol{\mathcal{D}}_\text{train}$, each pair $(\boldsymbol{\theta}_0,\boldsymbol{c})$ represents expert RIS phase shift $\boldsymbol{\theta}_0$ aligned with its corresponding channel state $\boldsymbol{c}$. This conditional modeling enables GCDM to generate optimal phase shift configurations under varying channel conditions as shown in Fig. \ref{fig_Gen_ai}.

The core concept of diffusion models is based on a  two-phase process: the forward process (noising) and the reverse process (denoising). In the forward process, expert determined RIS phase $\boldsymbol{\theta}_{0}$ are gradually distorted by Gaussian  noise with mean $m_t$ and variance $v_t$ until they become pure noise. Since noising process does not depend on channel state $\boldsymbol{c}$, we simplify $p(\boldsymbol{\theta_0}|\boldsymbol{c})$ to $p(\boldsymbol{\theta_0})$. In this forward process, noisy RIS phase shift \(\boldsymbol{\theta}_1, \boldsymbol{\theta}_2, \dots \boldsymbol{\theta}_T\) are obtained  through the following Markov chain with transition probability
\begin{equation}
p(\boldsymbol{\theta}_t| \boldsymbol{\theta}_{t-1}) \sim \mathcal{N}(\boldsymbol{\theta}_t; \sqrt{m_t} \boldsymbol{\theta}_{t-1}, v_t \boldsymbol{I}), \quad 1 \leq t \leq T,
\label{eq:forward_process1}
\end{equation}
where $T$ represents the number of diffusion steps, which is large enough to introduce noise gradually over time. However, obtaining noisy RIS phase  $\boldsymbol{\theta}_t$ at each step by fully relying on \eqref{eq:forward_process1} requires a large number of noising iterations. By exploiting the recursive nature of \eqref{eq:forward_process1} and taking $m_t = 1-v_t$, we directly sample RIS phase shift $\boldsymbol{\theta_t}$ using 
\begin{equation}
\boldsymbol{\theta}_t = \sqrt{\alpha_t}\boldsymbol{\theta}_{0} + \sqrt{(1-\alpha_t)}\boldsymbol{\epsilon},
\label{eq:forward_process3}
\end{equation}
where \( \alpha_t = \prod_{i=1}^{t} m_i\) and Gaussian noise $\boldsymbol{\epsilon} \sim \mathcal{N}(\boldsymbol{0, I})$.

In the reverse process, the model reconstructs the original optimal RIS phase theta by approximating the true posterior distribution through a parameterized Gaussian model, expressed as
\begin{equation}
q(\boldsymbol{\theta}_{t-1}| \boldsymbol{\theta}_{t}, \boldsymbol{c}) \sim \mathcal{N}(\boldsymbol{\theta}_{t-1}; \boldsymbol{\mu}_w(\boldsymbol{\theta}_t, \boldsymbol{c}, t), \boldsymbol{\Sigma}_w(\boldsymbol{\theta}_t, \boldsymbol{c}, t)).
\label{eq:Reverse_process_NN}
\end{equation}
We condition the learning process on the channel state  $\boldsymbol{c}$ to ensure the model learns to generate optimal RIS phase shifts tailored to the current environment for cell-free mMIMO systems. This conditioning plays a critical role, as the environment directly shapes the learning dynamics. During training, we provide  channel state  $\boldsymbol{c}$ and uniformly sampled time steps $t$ in the  neural network, enabling model to learn effective denoising across different stages of the diffusion process. To effectively guide the generation of the optimal 
RIS phase shift $\boldsymbol{\theta}$ for the current environment, we also concatenate the channel state $\boldsymbol{c}$ with the neural network input. In \eqref{eq:Reverse_process_NN}, we set variance \(\boldsymbol{\Sigma}_w(\boldsymbol{\theta}_t,\boldsymbol{c}, t) = \sigma_t^2\boldsymbol{I}\), where \(\sigma_t^2 \) is fixed as \(\frac{1-\alpha_{t-1}}{1-\alpha_{t}} v_t\) ~\cite{Lit_GenAI_2020_DDPM}, and the mean is given by
\begin{equation}
\boldsymbol{\mu}_w(\boldsymbol{\theta}_t, \boldsymbol{c}, t) = \frac{1}{\sqrt{m_t}} \left(  \boldsymbol{\theta}_t - \frac{v_t}{\sqrt{1-\alpha_t}}\boldsymbol{\epsilon}_w(\boldsymbol{\theta}_t, \boldsymbol{c}, t) \right),
\label{eq:mean2_NN}
\end{equation} 
 where \(\boldsymbol{\epsilon}_w\) is learned noise by neural network. We find the optimal RIS phase shift \(\boldsymbol{\theta}_P \) at time step $t=1$ during the denoising step, using  \(\boldsymbol{\epsilon}_w\), as follows
 \begin{equation}
 \boldsymbol{\theta}_{t-1} = \boldsymbol{\mu}_w(\boldsymbol{\theta}_t, \boldsymbol{c}, t) + \sigma_t \boldsymbol{\epsilon_t}, \quad \boldsymbol{\epsilon_t} \sim \mathcal{N}(\boldsymbol{0, I})
 \label{eq:mean2_NN2}
\end{equation}  
 Here, the model aims to find the optimal RIS phase shift that maximizes the sum SE by minimizing the loss function. This loss function, designed to reduce the gap between the sum SE achieved by the expert algorithm and our GCDM respectively, is given by 
 \begin{equation}
\mathcal{L}_t (w)\!=\! \mathbb{E}_{\boldsymbol{\theta}_{P}\sim \boldsymbol{\epsilon}_w}[{\|\mathcal{R}_\text{sum}(\boldsymbol{\theta}_{P}, \boldsymbol{c})\!-\!\mathcal{R}_\text{sum}(\boldsymbol{\theta}_{0}, \boldsymbol{c})\|}^2],
\label{eq:loss_fun_sumSE}
\end{equation}
where \(\boldsymbol{\theta}_P \) represents the RIS phase shifts predicted by GCDM. The loss function in \eqref{eq:loss_fun_sumSE} is simplified as 
\begin{equation}
\mathcal{L}_t (w)= \mathbb{E}[{\|\boldsymbol{\epsilon}-\boldsymbol{\epsilon}_w(\sqrt{\alpha_t}\boldsymbol{\theta}_{0} + \sqrt{(1-\alpha_t)}\boldsymbol{\epsilon}, t, \boldsymbol{c})\|}^2].
\label{eq:loss_fun_final}
\end{equation}
Therefore, by following the reverse diffusion  process and using a well-trained noise prediction neural network, we generate RIS phase shifts that closely match the distribution of the expert RIS phase shift from training dataset \( \boldsymbol{\mathcal{D}}_\text{train} \). The detailed steps for training  and sampling process of GCDM are presented in Algorithm~\ref{Alg_1}.

\begin{algorithm}
\caption{GCDM Based RIS Phase Shift Optimization}
\label{Alg_1}
\begin{algorithmic}[1]
  \Statex \hspace*{-0.5 cm} \textbf{Training Stage:}  
  \State Prepare an expert dataset~$\boldsymbol{\mathcal{D}}_\text{train}$ by generating optimal RIS phase shifts $ (\boldsymbol{\theta}_{0})$ using an expert algorithm for different CSI $(\boldsymbol{c})$.
  \State Initialize the number of epochs $E_p$, denoising steps $T$, training samples $N_T$, and GCDM parameters $w$.
  \For{$n \gets 1$ \textbf{to} $E_p$}
    \State Randomly shuffle samples of~$\boldsymbol{\mathcal{D}}_\text{train}$. 
    \For{$i \gets 1$ \textbf{to} $N_T$}
        \State Take $ (\boldsymbol{\theta}_{0}^{(i)}, \boldsymbol{c}^{(i)}) $ from shuffled $\boldsymbol{\mathcal{D}}_\text{train}$.
        \State Generate random gaussian noise $\boldsymbol{\epsilon} \sim \mathcal{N}(\boldsymbol{0, I})$. 
        \State Choose timestep $t \sim \text{Uniform}(1, 2, \dots T)$. 
        \State Get noisy RIS phase shift
        $\boldsymbol{\theta}_t^{(i)} $ using \eqref{eq:forward_process3}  by adding \Statex \hspace*{0.95 cm} noise $\boldsymbol{\epsilon}$ to expert RIS phase shift $\boldsymbol{\theta}_{0}^{(i)}$.
        \State Using $\boldsymbol{\theta}_t^{(i)}$, $\boldsymbol{c}$ and $t$, predict \(\boldsymbol{\epsilon}_w\) by neural network.
        \State Update the neural network using the loss function \Statex \hspace*{0.95 cm} defined in \eqref{eq:loss_fun_final}. 
    \EndFor
\EndFor
 
\State Return the trained model.
\end{algorithmic}

\vspace{0.5em}
\hrule

\begin{algorithmic}[1]
  \Statex \hspace*{-0.5 cm} \textbf{Sampling Stage:}  
    \State Get the current channel state $\boldsymbol{c}$.
    \State Generate random Gaussian $\boldsymbol{\theta}_T \sim \mathcal{N}(\boldsymbol{0, I})$. 
    \State Denoise $\boldsymbol{\theta}_T$ using trained model iteratively as follows:  
    \For{$t \gets T$ \textbf{to} $1$}
        \State Find $\boldsymbol{\theta}_{t-1} $ from $\boldsymbol{\theta}_t$ using \eqref{eq:mean2_NN2}.  
    \EndFor
\State Return the optimal RIS phase shift $\boldsymbol{\theta}_0$.

\end{algorithmic}
\end{algorithm}

\subsection{{GCDIM Based RIS Phase Optimization}}
GCDM involves executing the full sequence of reverse diffusion steps across the entire Markov chain, requiring step-by-step data generation from \(t = T\) to \(t = 0\). This sequential dependency creates a bottleneck in resource-intensive system and impractical for communication scenarios that demand low latency or operate under constrained computational resources. To address this limitation, we leverage denoising diffusion implicit model ~\cite{Lit_GenAI_2021_DDIM}, which reformulates the data generation process using a non-Markovian process and significantly enhances the sampling efficiency. We sample \(\boldsymbol{\theta}_{t-1}\) from a sample \(\boldsymbol{\theta}_{t}\) as follows
\begin{equation}
\boldsymbol{\theta}_{t-1} \!=\! \frac{1}{\sqrt{m_t}}\left(\boldsymbol{\theta}_{t} \!-\!(\sqrt{1-\alpha_{t}})(1\!-\!\sqrt{m_t})\boldsymbol{\epsilon}_w(\boldsymbol{\theta}_t, \boldsymbol{c}, t) \right).
\label{eq:DDIM}
\end{equation}
Using \eqref{eq:DDIM}, GCDIM can predict \(\boldsymbol{\theta}_{0}\) from \(\boldsymbol{\theta}_{T}\) with fewer steps (subset of \(T\)) compared to GCDM. Additionally, GCDIM employs the same training process as GCDM. The sampling process for GCDIM is explained in Algorithm~\ref{Alg_2}.

\begin{algorithm}
\caption{GCDIM Based RIS Phase Shift Optimization}
\label{Alg_2}
\begin{algorithmic}[1]
\Statex \hspace*{-0.5 cm} \textbf{Sampling Stage:}
    \State Get the current channel state $\boldsymbol{c}$.
    \State Generate random Gaussian $\boldsymbol{\theta}_T \sim \mathcal{N}(\boldsymbol{0, I})$. 
    \State Generate sub-steps, $ (\tau_1, \tau_2 \dots \tau_S) \subset (1,2, \dots T)$  with $S<<T$.
    \State Denoise $\boldsymbol{\theta}_T$ using trained model iteratively as follows:  
    \For{{$t \gets \tau_S $ \textbf{to} $\tau_1$}}
        \State Find $\boldsymbol{\theta}_{t-1} $ from $\boldsymbol{\theta}_t$ using \eqref{eq:DDIM}.
    \EndFor
    \State Return the optimal RIS phase shift $\boldsymbol{\theta}_0$. 

\end{algorithmic}
\end{algorithm}

\begin{figure*}[htbp]
	\centering
	\begin{subfigure}[b]{0.30\linewidth}
		\hspace{-25pt}\includegraphics[width=1.2\linewidth,height=1\linewidth]{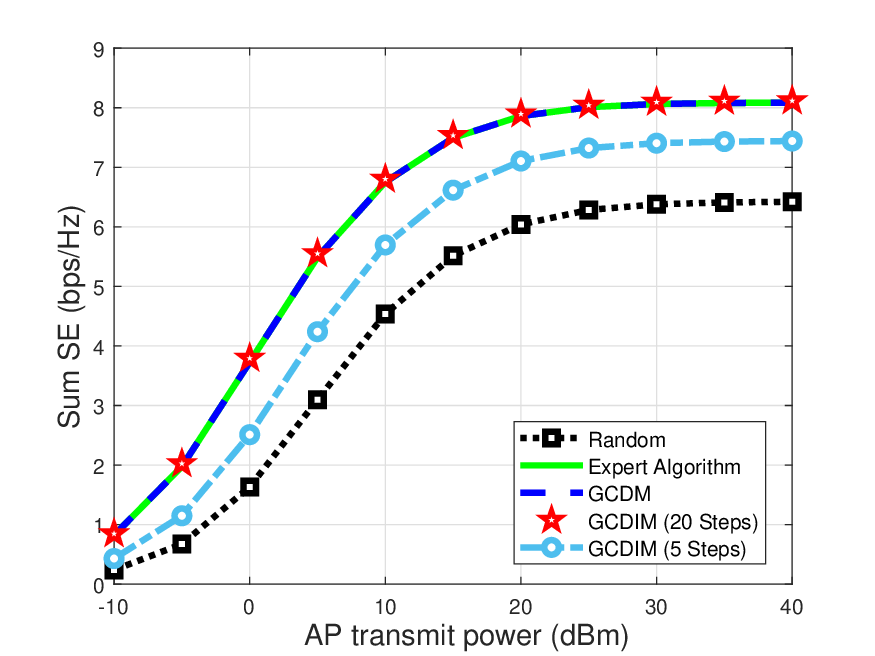} 
		\caption{\small} 
		\label{fig:1}
	\end{subfigure}\vspace{-2pt}
	\begin{subfigure}[b]{0.31\linewidth}
    
    \hspace{-12pt}\includegraphics[width=1.2\linewidth,height=0.97\linewidth]{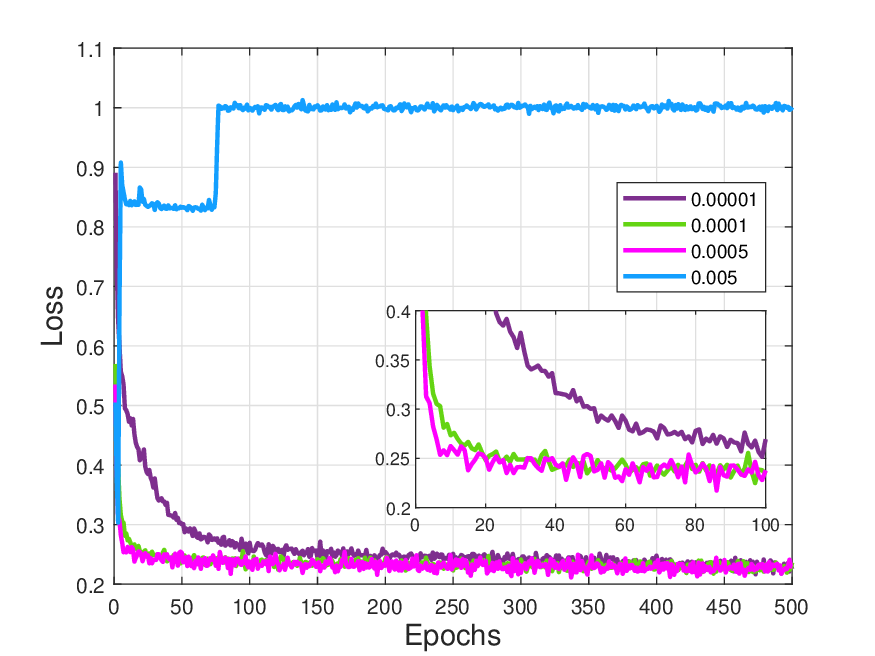}
	\caption{\small} 
	\label{fig:2}
	\end{subfigure}\vspace{-2pt}
	\begin{subfigure}[b]{0.30\linewidth}
    
	\hspace{2pt}\includegraphics[width=1.2\linewidth,height=1\linewidth]{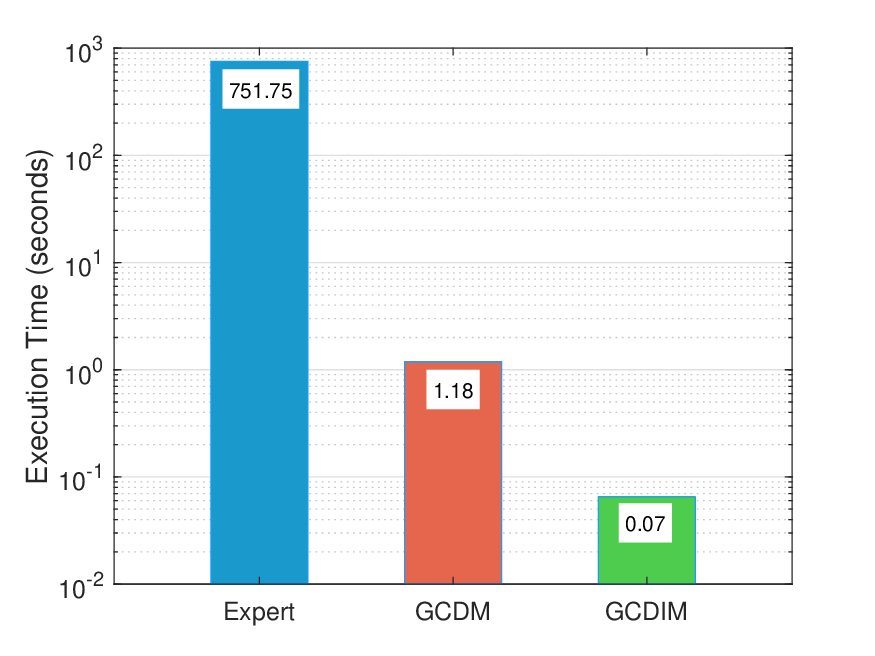}
	\caption{\small} 
	\label{fig:3}
	\end{subfigure} \vspace{-2pt}
	\caption{\small a) Sum SE versus AP transmit power, b) Training loss convergence for different learning rates, c) Comparison of the execution time across different algorithms. \\[-20pt]} 
\end{figure*} 

\section{{Numerical Results and Analysis}}
In this section, we numerically validate the performance of the GenAI based algorithms, i.e., GCDM and GCDIM, for the RIS-aided cell-free mMIMO system. For simulation, we consider a large geographical area of $1\,\text{km} \times 1\,\text{km}$, where all the APs and users are randomly distributed. A three-slope path-loss model is employed to determine the distance-dependent large-scale path loss~\cite{Lit_cellfree_2017}. We set $M=64$ APs, $K=12$ users, and $N=64$ RIS elements.
We use modified UNet architecture for noise prediction within the GenAI based model. The network comprises a single down-sampling layer in the encoder and a single up-sampling layer in the decoder, enhanced by one self-attention layers to improve performance. The input includes the optimal RIS phase shifts obtained from an expert dataset generated through a genetic algorithm~\cite{GA_1994}. The model also receives the noising time step $t$ and CSI $\boldsymbol{c}$ to incorporate conditional information. We train the model using a dataset containing different $\boldsymbol{c}$ values corresponding to transmit power levels ranging from $-10$ dB to $40$ dB. In GCDM, we set the diffusion steps $T=1000$ for training and sampling. In contrast, GCDIM uses the same trained network by reducing the sampling steps to $20$. The training is performed over $500$ epochs with a batch size of $8$, using the Adam optimizer with an adaptive learning rate initially set to $0.0005$. We conducted all the experiments on a Windows 11 system with a 10th Gen Intel(R) Core(TM) i7-10700 CPU @ 2.90 GHz and 16 GB RAM. 

Fig.~\ref{fig:1} illustrates a comparative analysis of three distinct RIS phase shift design strategies, i.e., unoptimized (Random), expert algorithm, and the proposed GenAI based approaches (GCDM and GCDIM), in terms of their impact on the sum SE in RIS-aided cell-free mMIMO systems. The proposed GenAI based approaches GCDM and GCDIM, consistently outperform the random phase shift configuration, highlighting the importance of optimized phase shift design. GCDM achieves the highest sum SE, closely matching the benchmark algorithms. Notably, GCDIM achieves performance comparable to GCDM with only 20 steps and maintains a favorable trade-off even when reduced to 5 steps. The ability of GCDIM to maintain similar results with fewer time steps can be attributed to its deterministic nature, which contrasts with the stochastic sampling process of GCDM, where randomness is inherently introduced. GCDIM proves practically effective by achieving near-optimal performance with significantly fewer diffusion steps. This confirms that GenAI based approaches can achieve performance comparable to traditional optimization methods, making them well suited for real time wireless communication scenarios.

Fig.~\ref{fig:2} shows the convergence of the training loss from \eqref{eq:loss_fun_final} for various learning rates over $500$ epochs. A learning rate of $0.005$ causes instability and divergence due to overly large updates. In contrast, learning rates of $0.0001$ and $0.0005$ yield stable convergence, with $0.0005$ offering faster and smoother loss reduction. A smaller rate of $0.00001$ converges slowly. The zoomed-in view of the first 100 epochs shows that a learning rate of 0.0005 achieves the best balance between convergence speed and stability.

Fig.~\ref{fig:3} compares the execution time of three different algorithms, i.e., expert algorithm, GCDM, and  GCDIM, highlighting their computational efficiency. The expert algorithm exhibits the highest execution time by a significant margin, taking approximately $752$ seconds, which reflects the high computational complexity associated with the iterative population-based optimization method.
In comparison, our GenAI based GCDM and  GCDIM, drastically reduce the execution time to around $1.18$ seconds and $0.07$ seconds, respectively, by leveraging a trained generative model, showcasing their practical advantage in real-time applications. We observe that GCDIM achieves the fastest execution time of just $0.07$ seconds due to its deterministic sampling mechanism that requires fewer diffusion steps. 

\section{Conclusion}
This work proposed two GenAI based approaches, GCDM and GCDIM, for optimizing the phase shifts in RIS-aided cell-free mMIMO systems, accounting imperfect CSI and spatial correlation. Both GCDM and GCDIM leverage diffusion models conditioned on dynamic CSI to maximize the sum SE of the system. Our results demonstrated that proposed GCDM achieved similar SE performance than the expert algorithm, while significantly reducing the time complexity. Furthermore, GCDIM offers a comparable SE with an additional $98\%$ reduction in computation time, highlighting its potential in optimizing the phase shifts in RIS-aided cell-free~systems.

\bibliographystyle{IEEEtran}
\bibliography{cite}

\end{document}